\documentclass[12pt]{article}

\newcommand{\textlineskip}{\baselineskip=13pt}

\def\lim{\rightarrow}
\def\half{\frac{1}{2}}
\def\quart{\frac{1}{4}}
\def\metl{g_{ij}}

\def\metu{g^{ij}}

\def\dphi{\dot{\phi}}
\def\dpsi{\dot{\psi}}

\def\apr{a^{\prime}(u)}
\def\bpr{b^{\prime}(v)}
\def\fid{\dot{\phi}}

\def\psid{\dot{\psi}}

\def\hd{\dot{h}}

\def\be{\begin{equation}}
\def\ee{\end{equation}}
\def\bea{\begin{eqnarray}}
\def\eea{\end{eqnarray}}

\def\dif{\partial}

\begin{document}
\title{Horizons in 1+1 Dimensional Dilaton Gravity Coupled to Matter}
\author{A.T.~Filippov\thanks{ filippov@thsun1.jinr.ru}\\
{\small \it  Joint Institute for Nuclear Research, Dubna, Russia}
\and   
D.~Maison\thanks{dim@mppmu.mpg.de}\\
{\small \it Max Planck Institut f\"ur Physik}\\
{\small \it Werner - Heisenberg Institut, Munich, Germany}
}

\date{}  % Deleting this command produces today's date.

\maketitle

\begin{center}
\bf{Abstract}
\end{center}

\bigskip

{\small
\noindent
We study the local behaviour of static solutions of a general 1+1 dimensional 
dilaton gravity theory coupled to scalar fields and Abelian gauge fields 
near horizons.
This type of model includes in particular reductions of higher 
dimensional theories invariant under a sufficiently large isometry group.
The solution near the horizon can in general be obtained solving a 
system of integral equations or in favourable cases in the form of a 
convergent series in the dilaton field.
}

\vspace*{0.3cm}\textlineskip
%\section{Introduction}
%\vspace*{-0.5pt}

\noindent
\noindent
{\bf 1.}~In the last decade 1+1 dimensional models of dilaton
gravity coupled to scalar matter fields were found to describe
some important properties of higher-dimensional black holes\footnote{
The best studied example is the string inspired dilaton gravity of
CGHS \cite{CGHS}. For a review see \cite{S}.}.
The connection between high and low
dimensions has been exploited in different contexts of gravity
and string theory  -  symmetry reduction, compactification,
holographic principle, AdS/CFT correspondence, duality, etc.\ 
(see e.g. \cite{BMG} -- \cite{Mo} and references therein).
An important case is given by spherically symmetric 
solutions of higher dimensional theories.
In dimension 1+1, these models are usually not integrable, but 
a generalization of the CGHS and Jackiw -- Teitelboim integrable 
models was introduced in \cite{A1}.
For its application to describing black hole evolution see e.g. \cite{Nav}.
The most general class of integrable models of dilaton gravity coupled to
scalar fields was recently proposed in \cite{A2}. 
A further reduction from 1+1 to 0+1 dimension can be performed for
time-independent solutions.
The resulting 0+1 dimensional dilaton gravity models are
frequently integrable in closed form 
(see e.g. \cite{VDA}, \cite{A1}, \cite{Kiem}),
although generically this is not the case.
For example, static, spherical black holes carrying Abelian gauge fields 
are described by integrable models, 
while non-Abelian gauge fields \cite{M1}
generically lead to non-integrable models (for an exception see
\cite{Volkov}).

In general the considered 1+1 dimensional models have no 
time-independent globally regular solutions, but they have static black hole 
solutions.
The most important properties of black holes (e.g. their mass, charge and 
their thermodynamics)
are encoded in the structure of their horizon. 
For this reason, local properties of
black hole horizons are of significant physical interest and merit
some special investigation.
Although in principle horizons are global concepts, not completely accessible
by local considerations, this is not the case for Killing horizons of
static black holes and this is what we shall consider henceforth.
We will show how one can find the analytic behaviour of the geometry and 
of the matter fields near the horizon for the class of models considered.
This does not require integrability of the equations, however
the results are valid only locally.

\noindent
{\bf 2.}~We consider a fairly general DG described by the Lagrangian density
\cite{A1}
\be
{\cal L} = \sqrt{-g} [U R(g) + V + W \metu \phi_i \phi_j
+ \sum^{K}_{k=1} X^{(k)} F^{(k)}_{ij} F^{(k)ij} + Y
+ \sum^{N}_{n=1} Z^{(n)} \metu \psi^{(n)}_i \psi^{(n)}_j] \; ,
\label{1}
\ee
where $F^{(k)}_{ij} = \dif_i A^{(k)}_j -  \dif_j A^{(k)}_i$,
$\phi_i = \dif_i \phi$, $\psi_i = \dif_i \psi$, $i,j = 0,1$.  $U, V, W$
are arbitrary functions of $\phi$ and $X, Y, Z$ -- arbitrary functions
of $\phi$ and $\psi$.
As is well known (and has been very simply and explicitly demonstrated
in Refs. \cite{A1}), the pure dilaton gravity coupled to Abelian
gauge fields $F_{ij}$ (when $\dif_{\psi}X \equiv 0$, $Y \equiv Z \equiv 0$)
is explicitly integrable for arbitrary functions $U(\phi)$, $V(\phi)$,
$W(\phi)$, $X(\phi)$. The general solutions are described by functions
of one variable.

It is convenient to write the solutions in the conformal metric
\be
ds^2 = -4f(u,v) du dv \, .
\ee

\noindent Then for the static solution we have
\be
\phi(u,v) = \phi(\tau) , \;\; f(u,v) = h(\tau) \apr \bpr , \;\;
\tau \equiv a(u) + b(v) ,
\label{2}
\ee
i.e. the metric is also defined by the function of one variable $\tau$,
up to conformal transformations $u \mapsto A(u)$, $b \mapsto B(v)$.
To further simplify the formulae we assume $U(\phi)\equiv\phi$.
achievable by a suitable field redefinition in a domain, where $U$ is
monotonous.
In addition we apply a Weyl transformation to make $W(\phi)$ zero,
requiring $U$ (resp.\ $\phi$) to be nonzero.
Then \cite{A1} one finds
\be
h(\tau) = M - N(\phi) ,
\label{3}
\ee
where
\footnote{Note that the equation of motion for $F^{(k)}_{ij}$
can be solved, because in dimension 1+1 the Maxwell fields do not
propagate. Thus, their effect can be included in the effective
action as a charge dependent piece in the potentials $V(\phi)$ or
$Y(\phi, \psi)$.}
\be
N(\phi) = \int d\phi [V(\phi) + \sum_{k}{2Q_k^2 / X^{(k)}(\phi)}]
\label{4}
\ee
and the dependence  of $\phi$  on $\tau$  can be found from
\be
\int d\phi [M - N(\phi)]^{-1} = \tau - \tau_0 \ .
\label{5}
\ee
Here $M$, $Q_k$ and $\tau_0$ are integrals of motion and the solutions of
the gauge field equations have the form
$F^{(k)}_{ab} = 2Q_k h(\tau)/X^{(k)}\phi)$ (we consider only electric charges
$Q_k$, magnetic ones being treated analogously).

Note that the static solution represents the general solution of 
pure dilaton gravity in terms of one free massless field $\tau = a(u) + b(v)$.
So, in this case, we might reduce the 1+1 dimensional theory to 0+1
dimension, solve the equations for the variables $\metl(\tau)$,
$\phi(\tau)$, $A(\tau)$, and then take $\tau = a(u) + b(v)$
 with arbitrary functions $a(u)$, $b(v)$.

Also note that this general static solution has at least one
horizon, because the equation $M - N(\phi) = 0$ has always at least one
root $\phi_0$ for some value of $M$.

For some special functions $U(\phi)$, $V(\phi)$, $W(\phi)$ and
$X(\phi)$ the model describes the
spherical solution of the Einstein -- Maxwell theory in any space-time
dimension. Let the Lagrangian density in $d$ dimensions be
\be
{\cal L} = \sqrt{-G}[R(G) - \quart {\rm e}^{\alpha \psi} F^2_{[2]} -
\half (\dif\psi)^2] .
\label{6}
\ee
With $\alpha =0$, this is the Einstein-Maxwell theory minimally
coupled to a scalar field $\psi$. In string theory this represents a
consistent truncation of some $d$-dimensional supergravities (in this
case instead of the antisymmetric Maxwell field two-form $F_{[2]}$ one has
to introduce $n$-forms $F_{[n]}$, corresponding  to the gauge potentials
$A_{[n-1]}$, see e.g. \cite{St}). Two-dimensional dilaton-gravity
theories may be obtained from (\ref{6})  by different sorts of dimensional
reductions. For illustrative purposes, it is sufficient to mention
the spherical reduction, when the $d$-dimensional metric is
\be
ds^2 = \metl dx^i dx^j  + {\rm e}^{-4\nu\phi} d\Omega^2_n \, ,
\label{7}
\ee
where $\nu = 1/n$, $n = d-2$, $x^0 = t$, $x^1 = r$; $\metl$ and $\phi$
depend only on $t$ and $r$;  $d\Omega^2_n$ is the metric on the $n$-dimensional
sphere. By integrating out the angle
variables in the action (\ref{6}) we may get the 1+1 dimensional action (\ref{1}) with
\bea
U& = &{\rm e}^{2\phi} \, , \;\; V = n(n-1){\rm e}^{-2\phi + 4\nu\phi} \, , \;\;
W = 4(1 - \nu){\rm e}^{-2\phi} \, ,\nonumber \\
X& = &-\beta {\rm e}^{-2\phi + \alpha\psi} \, , \;\;\; Y = 0 \, , \;\;\;
Z = - \gamma {\rm e}^{-2\phi} \, ,
\label{8}
\eea
where $\beta$  and $\gamma$ are some positive constants depending on the dimension $d$.

Unlike the case of pure dilaton gravity, the general solution of the 1+1
dimensional theory with non-constant scalar matter field $\psi$ essentially
depends on two variables and even simple theories of type (\ref{8}) may
not be integrable. Actually, integrability of the 1+1 dimensional theories
is rather an exception than the rule (well-known integrable examples are CGHS,
Jackiw-Teitelboim, Liouville and bi-Liouville models (see e.g.
\cite{A1}). However, to find the static black hole solutions (and their
brane generalizations) one has only to solve the 0+1 dimensional
sector of the theory (\ref{1}) and
many of the 0+1 dimensional models emerging from higher dimensions by
reductions  are explicitly analytically integrable. For example
reductions of (\ref{6}) to 0+1 dimensions lead to integrable systems. 
However, to describe horizons locally it is not necessary to use any sort of
global integrability. Thus  in \cite{A1} it was shown that any
theory (\ref{1}) with the potentials $U$, $V$, $W$, $X$, $Y$, $Z$ depending only
on $\phi$ does not allow static solution with horizons for non-constant scalar
fields, provided that the potentials
satisfy some  mild restrictions (absence of zeroes and poles).

\noindent
{\bf 3.} In order to study the local behaviour near horizons 
let us first rewrite
the Lagrangian density for the 0+1 dimensional sector
of the theory (\ref{1}). As already said, we put $U(\phi)$ equal to $\phi$ assuming that
$U(\phi)$ has no zeroes and is monotonous and use the Weyl transformation
$\metl \mapsto \Omega (\phi)\metl$ to make $W = 0$. This transformation is regular,
if $U(\phi)$ has no zeroes.
Finally, as explained above, we include the gauge
fields $F^{(k)}_{ij}$  into the effective potential depending on $Q_k$.
Then the Lagrangian density (\ref{1}) may be rewritten as
\be
{\cal L} = \sqrt{-g} \, [\phi R(g) + Y(\phi,\psi) +
\sum^N_{n=1} Z^{(n)}(\phi,\psi)\metu \psi^{(n)}_i \psi^{(n)}_j ] \, .
\label{9}
\ee
For reasons of simplicity we consider only the case of a single
matter field $\psi$, the generalization to any number of scalar fields being
straightforward. The static Lagrangian density corresponding to (\ref{9}) may be
written in the form (see \cite{A1})
\be
{\cal{L}}_{st} = -{1\over {l(\tau)}}[{\phi}^{\prime} h^{\prime} h^{-1} +
Z {{\psi}^{\prime}}^2] + l(\tau) [h Y(\phi, \psi)] \, ,
\label{10}
\ee
where $l(\tau)$ is a Lagrangian multiplier and the prime denotes the
differentiation with respect to $\tau$.

To investigate the horizons it is convenient to use the
Hamiltonian formulation (although our `time' $\tau$ is in reality a space
coordinate).
Due to the constraint it is
sufficient to consider the five first-order equations ($\bar{Z} \equiv Z^{-1}$)
\bea
\dphi& = &\chi \;\; {\rm (a)} ; \;\;\;\;\; \dot{\chi} = -hY \;\; {\rm (b)} ;
\label{11} \\
\dpsi& = &\bar{Z}\eta \;\; {\rm (a)} ;  \;\;\;
\dot{\eta} = -\half (hY_{\psi} + \eta^2 \bar{Z}_{\psi}) \;\; {\rm (b)} ;
\label{12} \\
\hd& = &-{h \over \chi} (hY + \bar{Z} \eta^2) \, .
\label{13}
\eea
Here the equations (\ref{11}a) and (\ref{12}a) are the definitions of the momentum
variables $\chi$  and $\eta$, (\ref{13}) is the constraint and the dot denotes
now differentiation with respect to the new variable $T \equiv \int dt l(\tau)$.
The variable $T$ coincides with $\tau$ in the gauge $l=1$.

A horizon is characterized by the vanishing of $h$, while
$\phi$ and $\psi$ remain finite. Obviously this is impossible at any finite
value of $T$ with $\chi\neq 0$, because then $h$ would be identically zero.
On the other hand, if $\chi$ and $h$ vanish at some finite value of $T$,
then $\chi$ also vanishes identically. Therefore a horizon can only be obtained for
$T\to\pm\infty$. This means that a horizon corresponds to a fixed point of the
`dynamical system' (\ref{11}) - (\ref{13}). Hence $\chi$ and $\eta$ have to tend
to zero for $T\to\pm\infty$. In order to desingularize (\ref{13}) we put
$h=\chi H$ and $\eta=\chi G$ and use $\chi$ as independent variable to get

\bea
{d\phi\over{d\chi}}& = &-{1\over{HY}},\;\;{d\psi\over{d\chi}}
=-{\bar{Z}G\over{HY}}\label{14}\\
{d(\chi G)\over{d\chi}}& = &{Y_{\psi}\over 2Y} + 
         {G^2\bar{Z}_{\psi}\chi\over{2HY}}, \;\;
{dH\over{d\chi}}= {\bar{Z}G^2\over {Y}}.
\label{15}
\eea
In order to obtain a horizon at a finite value $\phi_0$ of $\phi$
we have to assume
that $\chi$ vanishes there. Obviously the system (\ref{14}) - (\ref{15})
is singular at $\chi=0$ and a for regular solution the dependent variables 
have to fulfil further conditions. First we have to
require $Y_0= Y(\phi_0,\psi_0)\neq 0$, where $\psi_0=\psi(0)$. Furthermore
for a non-degenrate horizon we need $H_0=H(0)\neq 0$.
We replace
the system (\ref{14}) - (\ref{15}) by a system of integral equations
\bea
\phi&=&\phi_0-\int_{0}^{\chi}{d\chi^{\prime}\over{HY}}\\
\psi&=&\psi_0-\int_{0}^{\chi}{\bar{Z}Gd\chi^{\prime}\over{HY}}\\
H&=&H_0+\int_{0}^{\chi}{\bar{Z}G^2\over Y}d\chi^{\prime}\\
G&=&{1\over\chi}\int_{0}^{\chi} \Biggl( \chi^{\prime}
{G^2 \bar{Z}_{\psi}\over{2HY}}+{Y_{\psi}\over{2Y}} \Biggr)d\chi^{\prime}
\eea
incorporating the regular behaviour at $\chi=0$.
Starting with $\phi=\phi_0, \, \psi=\psi_0, \,
H=H_0, \, G = G_0={Y_{\psi}(0)\over{2Y(0)}}$ we may iterate these equations to obtain a
solution defined in a suitably small neighbourhood of $\chi=0$ \cite{M1}.
As is easily seen the solution inherits
the smoothness properties of the functions $Z(\phi,\psi)$ resp.
$Y(\phi,\psi)$. Thus in case the latter are analytic, we obtain a solution
that is analytic at the horizon.

A somewhat different approach consists in 
constructing a power-series expansion in $\chi$
of the solution near the horizon. Using the dependent variables of
Eqs.~(\ref{11})-(\ref{13}) we write the expansion
\bea
\eta & = &\eta_0 \chi (1 + \eta_1 \chi + \eta_2 \chi^2 + \ldots) \, , \nonumber \\
h & = &h_0 \chi (1 + h_1\chi  + h_2 \chi^2 + \ldots) \, , \nonumber \\
\phi & =&\phi_0 + \phi_1 \chi + \phi_2 \chi^2 \ldots \, , \nonumber \\
\psi & = &\psi_0 + \psi_1 \chi + \psi_2 \chi^2 + \ldots \, .
\label{22}
\eea
Correspondingly $Y$ and $\bar{Z}$ have to be expanded in powers of $\chi$
\bea
Y& = &Y_0 (1 + Y_1 \chi + Y_2 \chi + \ldots) \, , \qquad
Y_{\psi} = Y^{\prime}_0(1 + Y^{\prime}_1 \chi + Y^{\prime}_2 \chi^2 + \ldots) \, ,
\nonumber \\
\bar{Z}& = &\bar{Z}_0 (1 + \bar{Z}_1 \chi + \bar{Z}_2 \chi + \ldots) \, ,
\qquad
\bar{Z}_{\psi} = {\bar{Z}_0}^{\prime}(1 + {\bar{Z}_1}^{\prime} \chi +
{\bar{Z}_2}^{\prime} \chi^2 + \ldots) \, ,
\label{23}
\eea
where $Y_0$, $Y_1^{\prime}$, $\bar{Z}_0$, $\bar{Z}^{\prime}_0$
depend only on $\phi_0$, $\psi_0$ while $Y_n$, $Y_n^{\prime}$, $\bar{Z}_n$,
$\bar{Z}^{\prime}_n$ depend on $\phi_m$, $\psi_m$ with $m \leq n$.
In particular $Y_0 = Y(\phi_0,\psi_0)$, $Y^{\prime}_0 = Y_{\psi}(\phi_0, \psi_0)$, etc.

Now, substituting the expansions (\ref{22}), (\ref{23}) into the equations  
\be
h^{\prime} = (hY + \eta^2 \bar{Z})(\chi Y)^{-1}\;,  \quad {\rm (a)} \quad
\eta^{\prime} = (hY_{\psi} + \eta^2 \bar{Z}_{\psi})(2hY)^{-1} \quad {\rm (b)}
\label{20}
\ee
\be
{\phi}^{\prime} = -\chi (hY)^{-1}\;, \quad {\rm (a)} \quad
{\psi}^{\prime} = -\eta \bar{Z} (hY)^{-1} \quad {\rm (b)} \, .
\label{21}
\ee
we find recurrence relations, which can be solved order by order.
The crucial observation is that $h_0$, $\phi_0$, $\psi_0$ are not constrained by these
relations and that $\eta_0$ is given by the zeroth order term in Eq.~(\ref{20}b)
\be
\eta_0 = Y^{\prime}_0/2Y_0 = Y_{\psi}(\phi_0,\psi_0)/2Y(\phi_0,\psi_0)  \, .
\label{24}
\ee
Then from Eqs.~(\ref{20}a), (\ref{21}) we find $h_1$, $\phi_1$ and $\psi_1$:
\be
h_1 = {\frac {\eta_0^2 \bar{Z}_0} {h_0 Y_0}}
= {\frac {Y^{\prime 2} \bar{Z}_0}{4h_0 Y^3_0}} \quad
\phi_1 = -{\frac {1} {h_0 Y_0}} \quad \psi_1 =
-{\frac {Y^{\prime}_0 \bar{Z}_0} {2h_0 Y^2_0}} \, .
\label{25}
\ee
It is easy to check that this allows us to find $\eta_1$ from (\ref{20}b) and then
$h_2$, $\phi_2$, $\psi_2$. In general, once we know $\eta_{m-1}$, $h_m$, $\phi_m$,
$\psi_m$ for $m \leq n$, we immediately find $\eta_n$ and then $h_{n+1}$,
$\phi_{n+1}$, $\psi_{n+1}$.

The convergence of the Taylor series in the analytic case can be inferred
from the integral equation approach.
%%or more directly using the so-called criterion of `majorants+.

The dependence of the obtained solution on $T$ may be found from
Eq.(\ref{11}a) which now gives
\be
\phi_1 \ln \chi + 2 \phi_2 \chi + {\frac {3} {2}}\phi_3 \chi^2 + \ldots
= T - \tau_0 \, .
\label{26}
\ee
This can be solved iteratively near the horizon. The zeroth
approximation is $\chi_0 = \exp{(T - T_0)/{\phi_1}}$ which is exponentially 
small for
$\phi_1 > 0$ if $T \lim -\infty$.
Also the corrections decay exponentially near the horizon
$$
\chi = {\rm e}^{(T - T_0)/\phi_1} (1 - 2{\frac {\phi_2} {\phi_1}}
{\rm e}^{(T - T_0)/\phi_1} + \ldots) \, .
$$

\noindent
{\bf 4.}~In the $0+1$ dimensional formulation the (Killing) horizon is reached
at a singular point corresponding to its (`unphysical') bifurcation surface
\cite{Boyer}. 
In order to see that the solutions we have obtained
really correspond to a regular horizon we have to approach it along
light-like geodesics. This can be achieved using the light-like 
coordinates $u$ resp.\ $v$ together with a suitable space-like coordinate
serving as an affine parameter for in- resp.\ outgoing light rays.
For that purpose we rewrite the metric in the form
\be
ds^2 = 4h(\hat\tau) du^2-2dud\hat\tau \quad {\rm resp.} \quad 
ds^2 = 4h(\hat\tau)dv^2+2dvd\hat\tau\;.
\ee
From Eq.~(\ref{11}b) we find $d\hat\tau=-2/Yd\chi$. Since $Y\to Y_0\neq 0$
as $\chi\to0$ we see that the regularity in $\chi$ in fact guarantees the
regularity in $\hat\tau$. In addition the metrical function $h$ has a simple
zero in $\hat\tau$ as required for a non-degenerate horizon.  

\noindent
{\bf 5.}~In summary, we have found the solution near the horizon for
a general dilaton gravity coupled to scalar matter. As one can see, the
main condition for the existence of the obtained solutions is
$Y(\phi_0, \psi_0) \neq 0$, which means that the horizon is
nondegenerate. The degenerate case can be solved by using a similar technique
and will be treated elsewhere. The obtained formulae allow one to derive
all geometric and physical characteristics of the horizon. Considering the
0+1 dimensional model as a reduction of certain higher dimensional theory one
can also some characteristics of the horizon in the original higher
dimensional theory. Note that our approach and results may also be applied to
deriving asymptotic behaviour of some nonintegrable cosmological models.
\vskip5mm
This investigation was partly supported by the Russian Foundation for Basic
Research (project no. 00-01-00299).

\bigskip
\bigskip

\centerline{Appendix}

\bigskip

Consider the simplest theory
$$
{\cal L} = -{\frac {1} {l}}(\dphi \hd h^{-1} + {\dpsi}^2) + l h 2g {\psi}^{n} \, .
\eqno({\rm A}.1)
$$
The Lagrange equations,
$$
\ddot{\psi} = ghn{\psi}^{n-1} \, , \;\;\; (\dot{h}/h)^{\cdot} = 0 \, ,
\;\;\; \ddot{\phi} = - 2gh\psi^n \, ,
\eqno({\rm A}.2)
$$
give $h = \exp(\alpha + \beta T)$ and can easily be solved for $n = 1,2$.\footnote{
For $n=2$ solutions of the equation for $\psi$ can be expressed
in terms of the Bessel functions of $T$.}

In the simplest case, $n=1$, solving the equations and the
constraint we have
$$
\psi = \psi_0 + c_1 T + g\beta^{-2} {\rm e}^{\alpha + \beta T} \, ,
$$
$$
\phi = \phi_0 +c_1^2 \beta^{-1} T  - 2gc_1\beta^{-2} {\rm e}^{\alpha + \beta T} +
2g\beta^{-2}(2c_1 \beta^{-1} - \psi_0){\rm e}^{\alpha + \beta T} -
g^2 {(2\beta^4)}^{-1}{\rm e}^{2(\alpha + \beta T)} \, .
\eqno({\rm A}.3)
$$
Thus, we have a horizon for $T \lim \infty$ if and only if $c_1=0$. Then the
expansion of the solution in powers of $h$ is finite:
$$
\chi = \fid = -2gh\beta^{-1} \psi_0 [1 + gh {(2\psi_0 \beta^2)}^{-1}] \, ,
$$
$$
\eta = -\psid = -gh\beta^{-1} \, , \;\;\; \psi = \psi_0 + gh \beta^{-2} \, ,
\eqno({\rm A}.4)
$$
$$
\phi = \phi_0 - 2gh\psi_0 \beta^{-2} - g^2 h^2 {(2\beta^4)}^{-1} \, .
$$
One can rewrite it as an expansion in powers of $\chi$ but then it will be
an infinite series, whose radius of convergence is $\chi_c = {\psi_0}^2$.

Thus, for $\psi_0^2 \lim 0$ the radius of convergence is vanishing.
If $\psi_0 =0$, we have a degenerate case in the following sense:
$\chi \sim h^2$ and $(\phi - \phi_0) \sim h^2$
This means that $h(\phi)$ has a square root singularity at $\phi = \phi_0$.
Such a singularity would require a stronger singularity in the potential
$V(\phi)$ ($V \sim |\phi - \phi_0|^{-\half}$)
for the pure dilaton gravity and this is unnatural,
while the degenerate horizon, $h \sim (\phi - \phi_0)^n$, where $n$ is a positive integer,
is possible for analytic potentials.
Note however that the black hole solution of the $n=1$ model (A.1) with  $\psi_0=0$
is exceptional also for dilaton gravity coupled to matter, because in this case
$Y_0=0$ and the expansion (\ref{22}) is impossible (see Eq.(\ref{24})).

Of course, we may use a similar expansion in powers of $h$ and thus
reproduce the exact solution. But finding conditions for the existence
of degenerate solutions requires using more detailed information on
the potentials $Y$, $Z$. For example, the model (A.1) with $n=2$
does not allow the degeneracy. In this case $\psi$ and $\phi - \phi_0$ are convergent
series in powers of $2gh/\beta^2$:
$$
\psi = \psi_0 \sum^{\infty}_{n=0} (2gh\beta^{-2})^n (n!)^{-2} \, ,
$$
and $\phi$ can be obtained by integrating Eq.(A.2).
 Thus, they are analytic in $h$, and $(\phi - \phi_0) \sim h$, $(\psi - \psi_0) \sim h$.

\end{document}